\documentclass[twocolumn,tighten,times,fleqn]{aastex631}
\usepackage{color}
\usepackage{multirow}
\usepackage{chngcntr}
\usepackage{mathtools}
\usepackage{amsmath}
\usepackage{amsfonts}
\usepackage{amssymb}
\usepackage[utf8]{inputenc}  
\usepackage{graphicx}        
\usepackage{perpage}         
\usepackage{bm}
\usepackage{hyperref}
\usepackage{url}
\MakePerPage{footnote}

\graphicspath{{./}{}}

\makeatletter

\newcommand{\Rmnum}[1]{\expandafter\@slowromancap\romannumeral #1@}
\makeatother

\shorttitle{Electronic density calibration}
\shortauthors{Rong et al.}

\begin{document}

\title{A DESI Calibration of the [O II]--[S II] Electron-density Offset in Integrated Star-forming Galaxies}

\correspondingauthor{Yu Rong}
\email{rongyua@ustc.edu.cn}

\author{Yu Rong$^*$}
\affiliation{Department of Astronomy, University of Science and Technology of China, Hefei, Anhui 230026, China}
\affiliation{School of Astronomy and Space Sciences, University of Science and Technology of China, Hefei 230026, Anhui, China}

\author{Shihong Liu}
\affiliation{Department of Astronomy, University of Science and Technology of China, Hefei, Anhui 230026, China}
\affiliation{School of Astronomy and Space Sciences, University of Science and Technology of China, Hefei 230026, Anhui, China}

\author{Hu Zou}
\affiliation{Key Laboratory of Optical Astronomy, National Astronomical Observatories, Chinese Academy of Sciences, Beijing 100012, China}
\affiliation{School of Astronomy and Space Science, University of Chinese Academy of Sciences, Beijing 101408, China}

\begin{abstract}
The [O II]$\lambda\lambda3726,3729$ and
[S II]$\lambda\lambda6716,6731$ doublets are widely used as low-ionization
electron-density diagnostics in galaxy spectra and are often treated as
interchangeable when only one of them is accessible. We test this assumption
using the DESI DR1 Emission Line Catalog. For star-forming galaxies with fiducial emission lines, [O II] yields systematically higher electron
densities than [S II], with a median offset of
0.228 dex.
The binned median calibration is
$\log n_e({\rm OII})=(0.752^{+0.182}_{-0.097})\log n_e({\rm SII})
+(0.832^{+0.231}_{-0.422})$. The offset is larger in galaxies with higher
stellar mass, H$\alpha$ star-formation rate, dust attenuation, and $N2\equiv\log([{\rm NII}]\lambda6583/{\rm H}\alpha)$, an empirical gas-phase metallicity proxy, and
smaller in galaxies with higher \(\log O_{32}\equiv\log\{[{\rm OIII}]\lambda5007/
([{\rm OII}]\lambda3726+\lambda3729)\}\), an ionization proxy; no significant trend is found
with specific star-formation rate. These trends are consistent with [O II] and
[S II] sampling different low-ionization gas phases in integrated spectra,
with [S II] more strongly weighted toward lower-density diffuse or outer gas.
Our results show that [O II]- and [S II]-based densities should not be mixed
without empirical calibration in studies of ISM pressure, nebular density, and
their evolution across galaxy samples and redshift.
\end{abstract}


\section{Introduction}

The electron density of ionized gas is a fundamental quantity in nebular
astrophysics. It enters the calculation of collisionally excited-line
emissivities, abundance determinations, ionization-parameter estimates,
gas-pressure measurements, and, more generally, the interpretation of how star
formation couples to the interstellar medium
\citep[e.g.,][]{osterbrock2006,proxauf2014,kewley2019}. In unresolved galaxy
spectra, the density is commonly inferred from forbidden-line doublets whose
upper levels have different collisional de-excitation rates. Among the most
widely used diagnostics are the optical
[S II]$\lambda6716/\lambda6731$ and [O II]$\lambda3729/\lambda3726$ ratios.

The importance of these diagnostics extends beyond the local Universe.
Much of the empirical evidence for elevated gas densities in high-redshift
star-forming galaxies is based on comparisons of samples for which different
doublets are accessible: [S II] is routinely used at low redshift, whereas
[O II] is often the more available or higher signal-to-noise ratio (S/N) density diagnostic for many
intermediate- and high-redshift datasets
\citep[e.g.,][]{steidel2014,sanders2016,kaasinen2017,strom2017,davies2021}.
If the two doublets are interchangeable, such comparisons are relatively
straightforward. If they are not, part of the inferred density evolution may
instead reflect a diagnostic-dependent systematic offset.

There are several reasons to question the assumption of interchangeability.
The [O II] and [S II] doublets arise from different ions, have different
critical densities and atomic parameters, and need not be emitted with the
same weight across classical H II regions, partially ionized boundary layers,
and diffuse ionized gas along the line of sight. Photoionization calculations
show explicitly that [O II] and [S II] sample different low-ionization zones
and should not be assumed to yield identical densities unless the temperature
and density structure are tightly controlled \citep{kewley2019}. High-S/N
studies of nearby H II regions likewise show that unresolved or stratified
nebulae can bias classical low-ionization density diagnostics
\citep{mendezdelgado2023}. Diffuse ionized gas is also known to enhance
low-ionization emission and to bias integrated nebular diagnostics in galaxy
spectra \citep{zhang2017,sanders2017,lacerda2018,belfiore2021}. These
considerations motivate a simple empirical question: in integrated spectra of
normal star-forming galaxies, what relation actually connects
$n_e({\rm OII})$ and $n_e({\rm SII})$?

The Dark Energy Spectroscopic Instrument (DESI) provides the wavelength
coverage, spectral resolution, and sample size needed to address this question
statistically. In this work we use the DESI Data Release 1 (DR1) Stellar Mass
and Emission Line Catalog 
\citep{zou24} to calibrate the $n_e({\rm OII})$ and $n_e({\rm SII})$ relation. 
This Letter is organized as follows. Section~2 describes the DESI catalog,
spectral-resolution considerations, the star-forming selection, and the
density measurements. Section~3 presents the empirical [O II]--[S II]
calibration and a set of robustness tests, including cuts in density
uncertainty, doublet-ratio range, observed wavelength, ionization state, and
temperature. Section~4 discusses the likely physical origin of the offset and
its implications for comparing electron densities across galaxy samples and
redshift.

\section{Data and Measurements}

\subsection{Sample}

We use the public DESI DR1\footnote{https://data.desi.lbl.gov/doc/releases/dr1/} \citep{desidr1} Stellar Mass and Emission Line Catalog\footnote{https://data.desi.lbl.gov/doc/releases/dr1/vac/stellar-mass-emline/} of
\citet{zou24}. The redshifts used in this work are the official DESI
spectroscopic pipeline redshifts. The value-added catalog of \citet{zou24}
uses those redshifts and performs stellar-population synthesis fitting with
\textsc{STARLIGHT}, emission-line measurements after continuum subtraction,
and photometric SED fitting for stellar masses and related galaxy properties.
It provides the emission-line fluxes and uncertainties used below.
Our analysis uses the measured fluxes of [O II]$\lambda3726$,
[O II]$\lambda3729$, H$\beta$, [O III]$\lambda5007$, H$\alpha$,
[N II]$\lambda6583$, [S II]$\lambda6716$, and [S II]$\lambda6731$.
Before forming line ratios that span a broad wavelength baseline, we correct
the emission-line fluxes for dust attenuation using the Balmer decrement. We
assume an intrinsic H$\alpha$/H$\beta$ ratio of 2.86 for case-B recombination
and adopt the \citet{cardelli1989} extinction curve with $R_V=3.1$, setting
negative inferred color excesses to zero. All line ratios that span a broad
wavelength baseline are computed from dust-corrected fluxes. We define
\(O_{32}\equiv[{\rm OIII}]\lambda5007/
([{\rm OII}]\lambda3726+[{\rm OII}]\lambda3729)\), which we use as a proxy for
the ionization state or ionization parameter of the nebular gas, and define
\(N2\equiv\log([{\rm NII}]\lambda6583/{\rm H}\alpha)\), which we use as an
empirical proxy for gas-phase metallicity. H$\alpha$-based star-formation rates
are computed from dust-corrected H$\alpha$ luminosities using the calibration
of \citet{kennicuttevans2012}.

The DESI spectra have sufficient resolving power for the doublet measurements
used here. In the blue arm, DESI provides
$R\equiv\lambda/\Delta\lambda\simeq2000$--3200 over 3600--5550 \AA, while the
full instrument reaches higher resolving power at redder wavelengths
\citep{desidr1,desispectrograph}. The [O II] doublet separation is 2.78 \AA\
in the rest frame and scales as $(1+z)$ in the observed frame, whereas the
[S II] doublet separation is 14.38 \AA. The [S II] doublet is therefore
cleanly resolved, and the [O II] doublet is at least partially resolved at
DESI blue-arm resolution, with the separation improving at higher redshift.

For the fiducial sample used below, the median redshift is $z=0.109$, placing
[O II] at a median observed wavelength of 4135 \AA. Interpolating across the
DESI blue-arm resolving-power range gives a characteristic
$R_{\rm DESI}\simeq2300$, corresponding to an instrumental FWHM of
$\simeq1.8$ \AA\ at this wavelength. By comparison, the legacy SDSS
spectrograph has $R\simeq1850$--2200 over 3800--9200 \AA\
\citep{sdssdr3spectro}; the SDSS spectroscopic overview also quotes
$R\simeq1500$ at 3800 \AA\ increasing to higher values at redder wavelengths
\citep{sdssspectrobasics}. At the same [O II] wavelengths, DESI therefore
improves the resolving power by roughly a factor of \(1.2\)–\(1.4\) relative
to SDSS. Equivalently, the median [O II] doublet separation is 3.09 \AA,
corresponding to separation/FWHM \(\simeq1.7\) for DESI but only
\(\simeq1.2\)–1.4 for SDSS. DESI therefore does not make [O II] a
high-resolution diagnostic in an absolute sense, and density estimates near
the low- or high-density limits of the doublet remain sensitive to small
flux-ratio errors. Nevertheless, DESI moves the [O II] doublet from a
marginally resolved SDSS regime into a more reliable partially resolved
regime. Since the catalog fits the two [O II] components as individual
emission lines, the DESI resolution is adequate for a statistical
line-ratio-based density analysis, provided that we impose the S/N,
ratio-range, and density-error cuts described below.

\begin{figure*}
\centering
\includegraphics[width=0.96\textwidth]{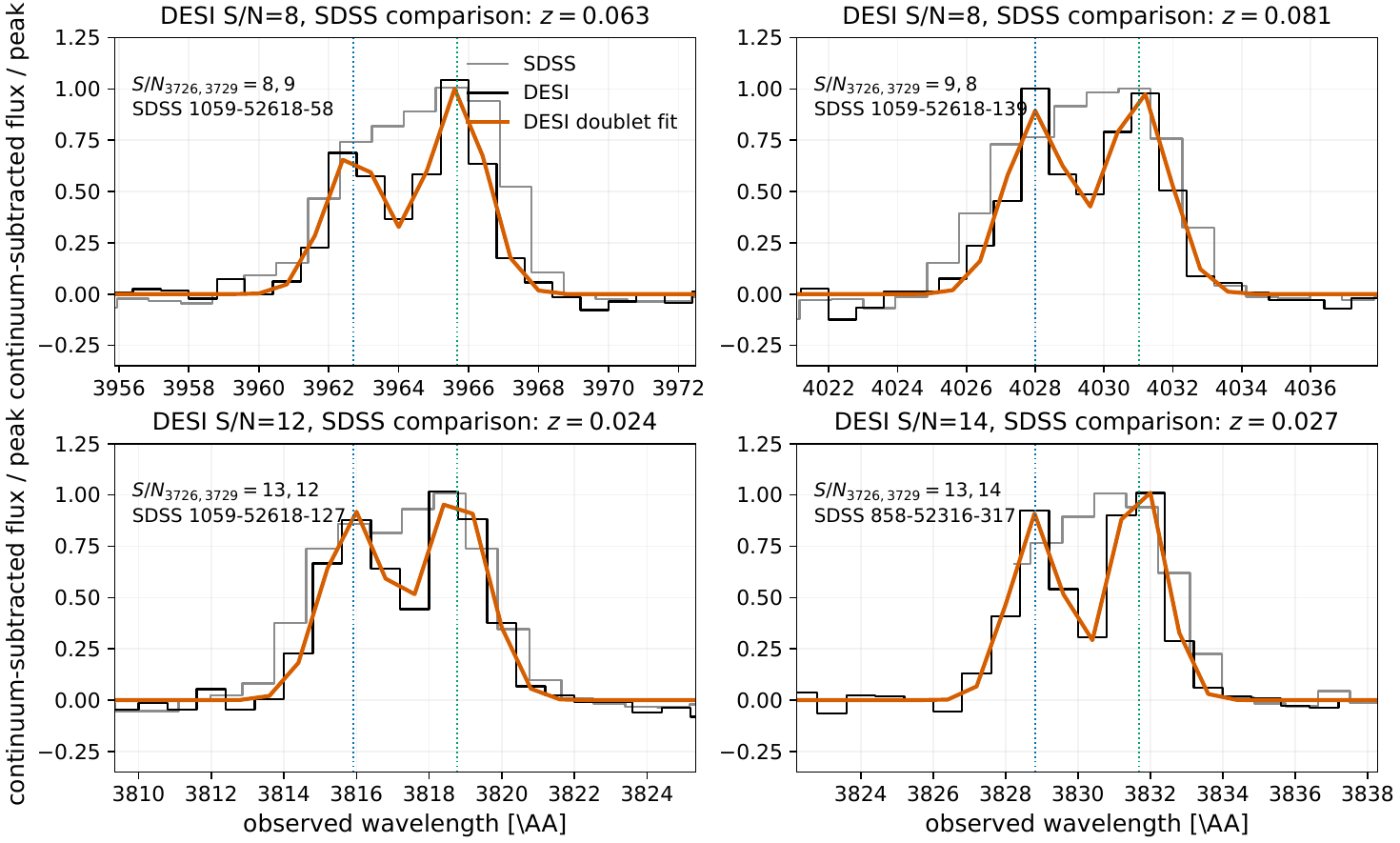}
\caption{Same-galaxy comparison of DESI and SDSS spectra around the
[O II]$\lambda\lambda3726,3729$ doublet. Black histograms show DESI coadd
spectra, gray histograms show SDSS spectra, and orange curves show simple DESI
local continuum-plus-doublet fits. For visual comparison only, each spectrum
is locally continuum-subtracted and normalized by its [O II] peak. Vertical
dotted lines indicate the observed-frame centers of the two [O II] components
from the local DESI doublet fit. The SDSS profiles are more blended, whereas
the DESI profiles show a clearer two-component structure at the S/N levels
relevant for the density analysis.}
\label{fig:oiiexamples}
\end{figure*}

Figure~\ref{fig:oiiexamples} shows four example galaxies which have both DESI
and SDSS spectra. For the same objects, the DESI coadds show a clearer
two-component [O II] profile than the legacy SDSS spectra, for which the
valley between the two components is far less distinct. 

We select galaxies with positive DESI pipeline redshifts and $z<0.5$, and require signal-to-noise ratio
S/N$>5$ in all lines entering the density and Baldwin-Phillips-Terlevich diagram \citep[BPT;][]{baldwin1981} selections:
[O II]$\lambda3726$,
[O II]$\lambda3729$, H$\beta$, [O III]$\lambda5007$, H$\alpha$,
[N II]$\lambda6583$, [S II]$\lambda6716$, and [S II]$\lambda6731$. We select
star-forming galaxies using the \citet{kauffmann2003} demarcation in the
[N II] BPT diagram
\citep{kewley2001}, requiring
$\log([{\rm NII}]\lambda6583/{\rm H}\alpha)<0$. Objects in the composite/AGN
region above the \citet{kauffmann2003} curve are excluded before any density
calibration is fit.

\subsection{Measurement}

Electron densities are computed with \textsc{PyNeb} \citep{pyneb}.
\textsc{PyNeb} solves the statistical-equilibrium equations for multi-level
ions using adopted transition probabilities and collision strengths, and
predicts line emissivities as functions of electron temperature and density.
For [S II] we invert the predicted ratio
$I(\lambda6716)/I(\lambda6731)$ to obtain $n_e({\rm SII})$; for [O II] we
invert $I(\lambda3729)/I(\lambda3726)$ to obtain $n_e({\rm OII})$. All
electron densities are number densities in cm$^{-3}$, and throughout this
paper $\log n_e$ denotes the base-10 logarithm of $n_e/{\rm cm}^{-3}$.

Our fiducial calculation adopts an electron temperature \(T_e=10^4\) K,
appropriate for a first empirical low-redshift star-forming-galaxy calibration
and standard in many strong-line applications \citep[e.g.,][]{osterbrock2006,proxauf2014,sanders2016}. Since each density diagnostic
is a close doublet, reddening corrections are negligible for the density ratios
themselves. Flux errors are propagated into doublet-ratio uncertainties and
then through the same \textsc{PyNeb} ratio--density grids to obtain the
\(1\sigma\) uncertainty \(\sigma(\log n_e)\). We later repeat the calculations
over \(T_e=7000\)–\(20000\) K to test whether the median offset is sensitive to
the fixed-temperature assumption.
\textsc{PyNeb} also provides joint temperature--density solvers, but such an
iteration requires a reliable temperature-sensitive auroral-line diagnostic,
for example [O III]\(\lambda4363\) or [N II]\(\lambda5755\), for the same
objects. Those auroral lines are much weaker than the density doublets in
integrated DESI galaxy spectra, so requiring them would reduce the present
high-precision density sample to a small and strongly auroral-line-selected
subset. Moreover, because the [O II] and [S II] density diagnostics are close
doublets, their inferred densities depend only weakly on plausible
temperature variations; this is quantified in Section~4.

The full BPT-selected, S/N$>5$, finite-density sample contains 103,548
galaxies. Our fiducial calibration subset additionally requires both [O II]-
and [S II]-based density uncertainties to be below 0.3 dex, leaving 1938
galaxies. We adopt this threshold as a compromise between precision and sample
size: a stricter 0.2 dex cut leaves only 162 objects, while no galaxy in the
selected catalog satisfies \(\sigma(\log n_e)<0.1\) dex for both diagnostics.
The full sample establishes the sign and ubiquity of the offset, whereas the
fiducial subset is used for fitting the calibration and for property-correlation
tests. We therefore do not interpret the fiducial subset as a complete
demographic census of all DESI star-forming galaxies.

\section{Results}

\begin{figure*}
\centering
\includegraphics[width=0.98\textwidth]{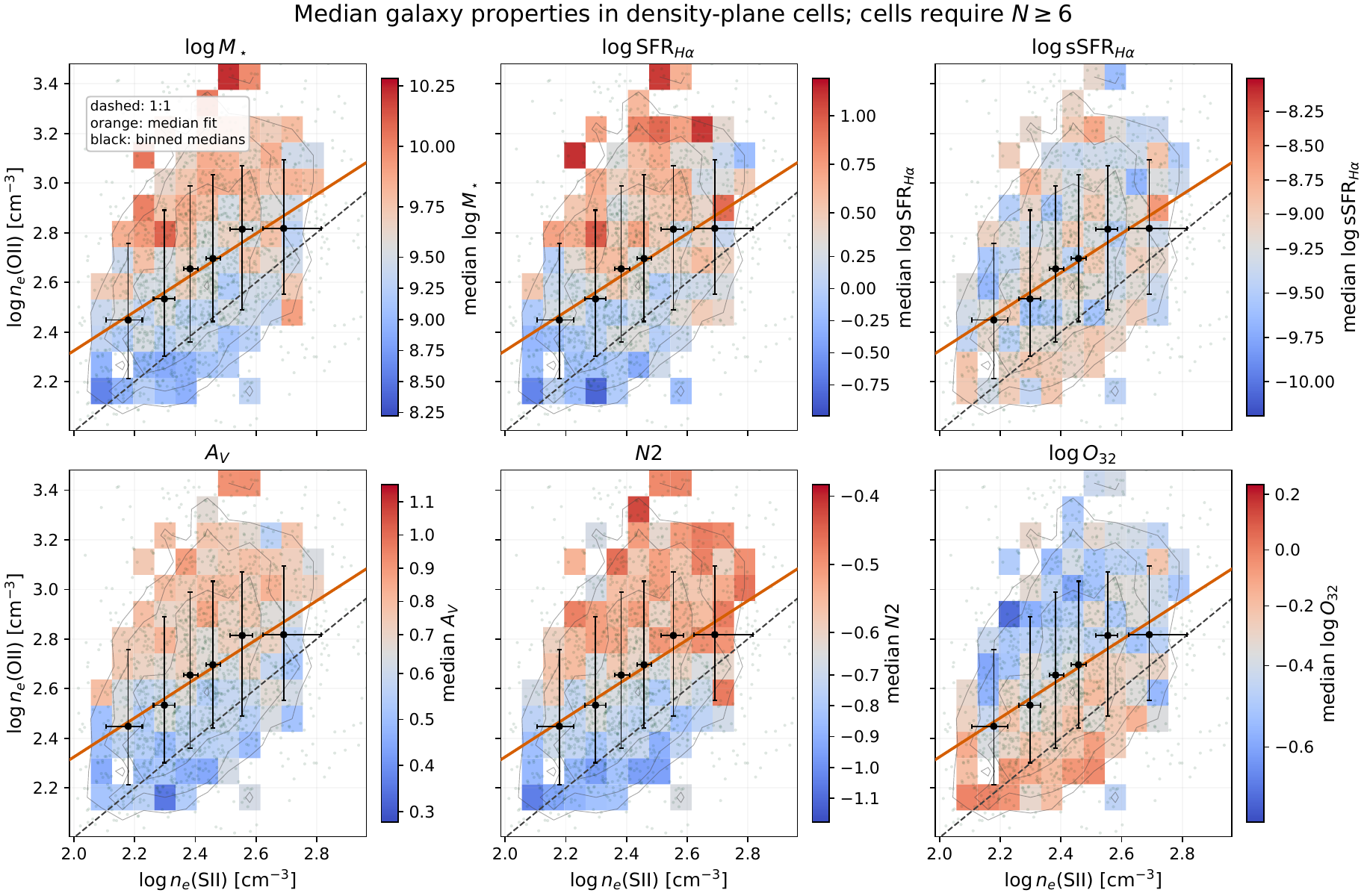}
\caption{The DESI [O II]--[S II] density plane for the fiducial
\(\sigma(\log n_e)<0.3\) dex subset. Each cell is colored by the median value
of the indicated galaxy property; green points show individual galaxies, black
points with error bars show the binned median density relation, and the orange
line is the median calibration fit. The black points are medians in
equal-number bins of \(\log n_e({\rm SII})\); the horizontal and vertical error
bars show the 16th--84th percentile ranges of \(\log n_e({\rm SII})\) and
\(\log n_e({\rm OII})\) within each bin. The dashed line marks equality. The
binned medians show that [O II] yields systematically higher densities than
[S II], while the cell colors reveal clear gradients across the density plane.
These gradients alone do not determine whether the relevant galaxy properties are primarily linked to the absolute electron density or to the [O II]--[S II] offset; Figures~\ref{fig:propertyfits} and \ref{fig:propertyresid} test these two possibilities explicitly.
}
\label{fig:relation}
\end{figure*}

\subsection{A Median [O II]--[S II] Density Calibration}

Figure~\ref{fig:relation} shows the fiducial
\(\sigma(\log n_e)<0.3\) dex subset. The galaxies are not symmetrically
distributed about the one-to-one relation. The median offset is
\begin{equation}
\Delta\log n_e =
\log n_e({\rm OII})-\log n_e({\rm SII})
\simeq 0.228~{\rm dex},
\end{equation}
and 79.6\% of the galaxies satisfy \(n_e({\rm OII})>n_e({\rm SII})\).

We fit the binned median relation using eight equal-number bins in
\(\log n_e({\rm SII})\).
The resulting empirical calibration is
\begin{equation}
\log n_e({\rm OII}) =
(0.752^{+0.182}_{-0.097})\,\log n_e({\rm SII})
+(0.832^{+0.231}_{-0.422}),
\label{eq:calibration}
\end{equation}
where the quoted uncertainties are obtained by bootstrapping the binned
medians. 
Thus the offset is largest at low [S II]-based density and decreases toward
higher density. At \(\log n_e({\rm SII})=2.2\), 2.5, and 2.7, the calibration
implies offsets of 0.29, 0.21, and 0.16 dex, respectively.

\subsection{Galaxy-Property Trends in the Density Plane}

\begin{figure*}
\centering
\includegraphics[width=0.98\textwidth]{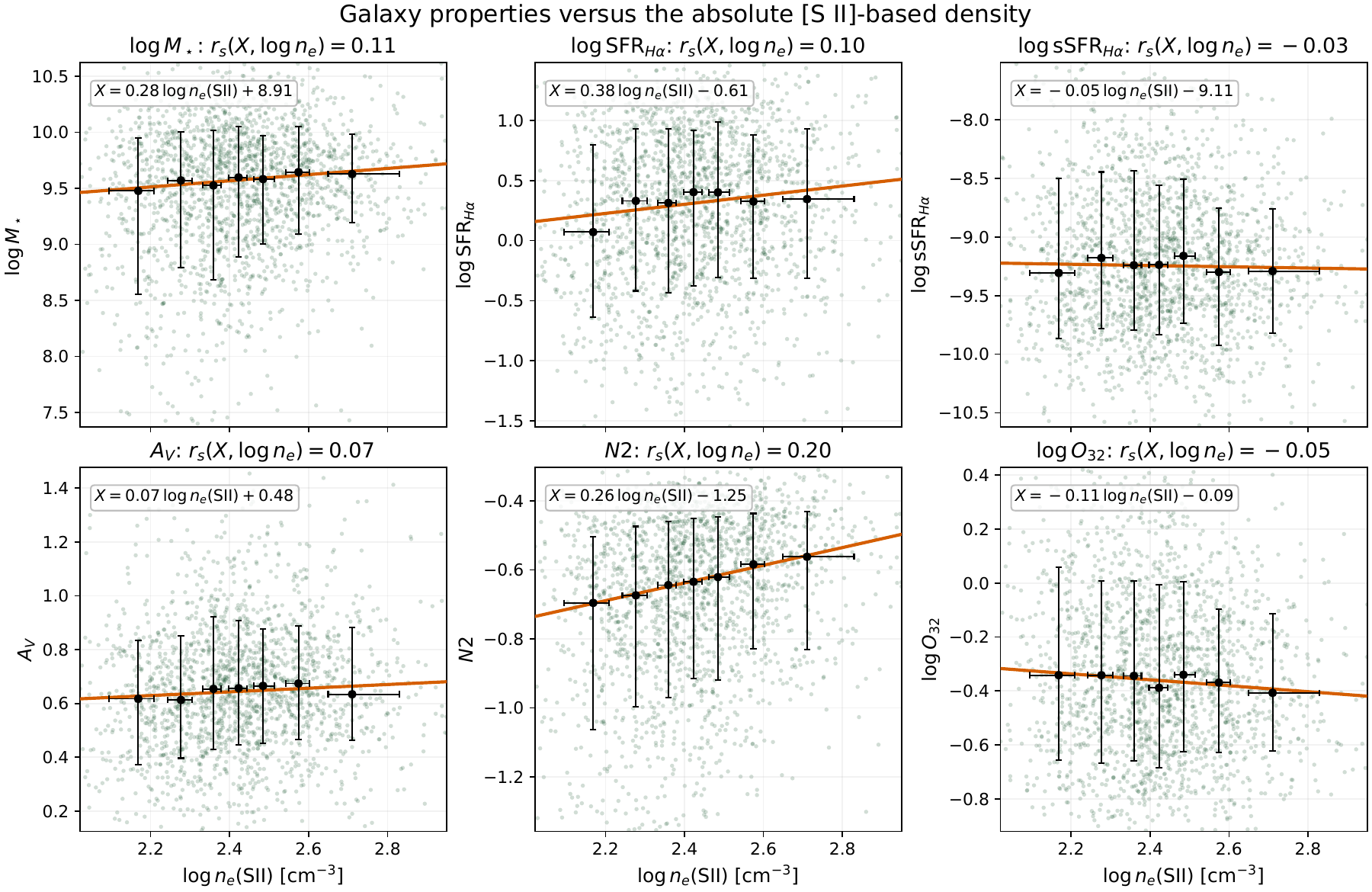}
\caption{Galaxy-property trends with the [S II]-based electron density,
shown as a control comparison for Figure~\ref{fig:propertyresid}. Small green
points show individual galaxies within the displayed plotting range, black
points with error bars show equal-number binned medians and central 68\%
intervals, and orange lines show linear fits to the binned medians as visual
guides. The fitted relation is written in each panel, with \(n_e\) in
cm$^{-3}$. The axes are clipped to the main data locus so that a small number
of outliers do not determine the visual scale; the Spearman rank correlation
coefficients, \(r_s\), are computed from the full fiducial sample.}
\label{fig:propertyfits}
\end{figure*}

Figure~\ref{fig:relation} also shows that galaxy properties are not randomly
distributed across the [O II]--[S II] density plane. Stellar mass,
SFR, dust attenuation, and \(N2\) tend to be larger in the regions
where [O II] gives a larger density than [S II], while \(\log O_{32}\) shows
the opposite sense. These color gradients are visually suggestive, but by
themselves they do not identify whether the relevant properties are tied to
the absolute density scale, to the diagnostic offset, or to both. We therefore
separate the problem into two tests below: first comparing each property with
\(\log n_e({\rm SII})\), and then comparing it directly with the offset
\(\Delta\log n_e=\log n_e({\rm OII})-\log n_e({\rm SII})\).

Figure~\ref{fig:propertyfits} tests the first possibility by plotting each
property against the absolute [S II]-based density. The trends are generally
weak. In particular, \(\log n_e({\rm SII})\) has only weak rank correlations
with dust-corrected H$\alpha$-based SFR (\(r_s=0.10\)) and \(\log O_{32}\)
(\(r_s=-0.05\)), where \(r_s\) denotes the Spearman rank correlation
coefficient; the corresponding median relations are shallow compared with
the dynamic range of the properties. Therefore, the color gradients in
Figure~\ref{fig:relation} are not simply a projection of strong
property--\(n_e({\rm SII})\) relations.

\begin{figure*}
\centering
\includegraphics[width=0.98\textwidth]{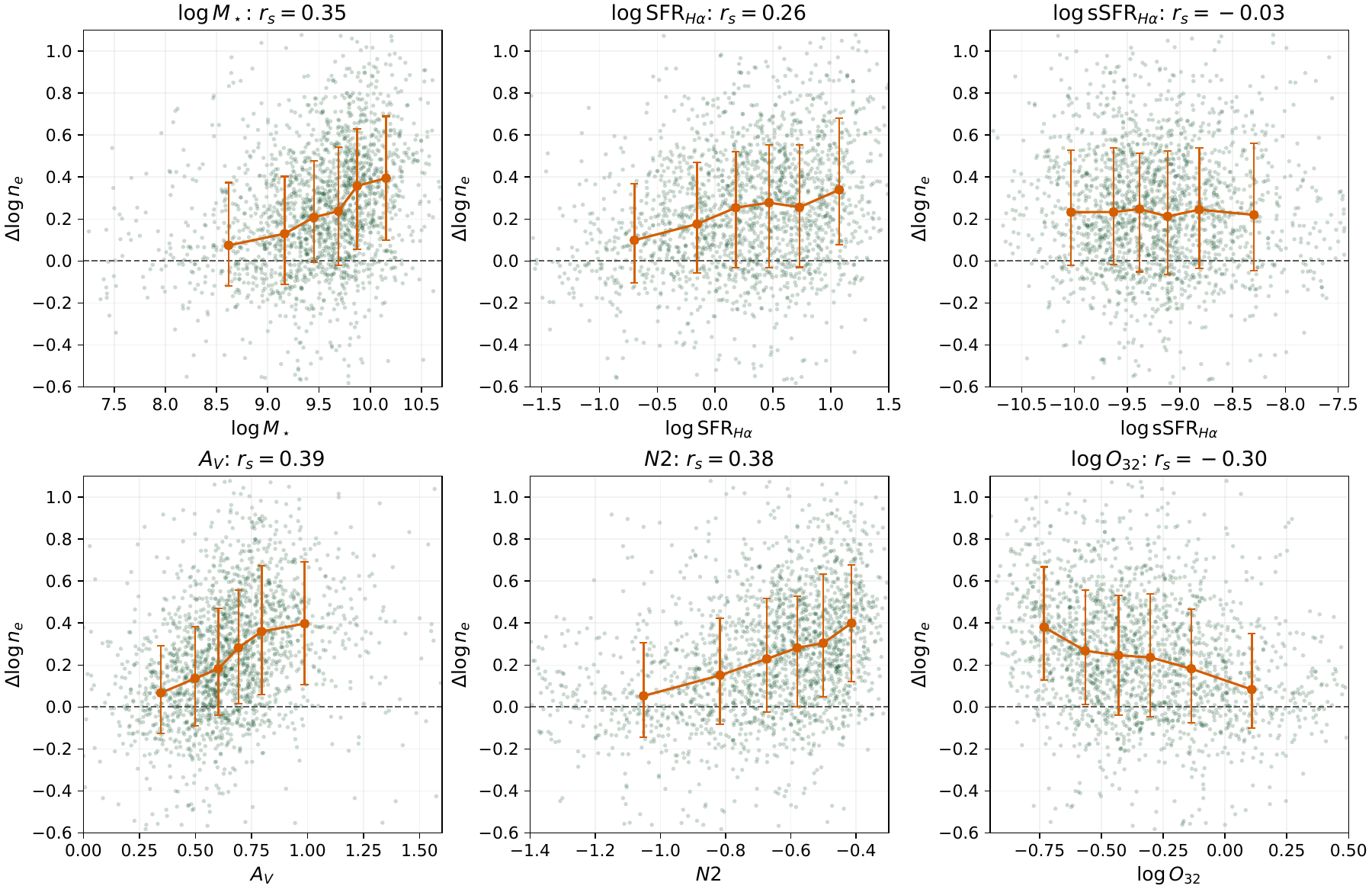}
\caption{Direct correlations between galaxy properties and the [O II]--[S II]
density offset. For each property \(X\), we plot \(X\) against
\(\Delta\log n_e=\log n_e({\rm OII})-\log n_e({\rm SII})\).
Small green points show individual galaxies within the displayed plotting
range; orange points and curves show equal-number binned medians. The title of
each panel gives the Spearman rank correlation coefficient, \(r_s\), computed from the full fiducial
sample. Stellar mass, H$\alpha$ SFR, dust attenuation, \(N2\), and
\(\log O_{32}\) correlate with the diagnostic offset, whereas sSFR does not.
Comparison with Figure~\ref{fig:propertyfits} shows that these properties
primarily track the offset between diagnostics rather than the absolute
[S II]-based density.}
\label{fig:propertyresid}
\end{figure*}

Figure~\ref{fig:propertyfits} provides the necessary control comparison: most
of the plotted galaxy properties are weak predictors of the absolute [S II]-based
density. Figure~\ref{fig:propertyresid} then gives the complementary direct
test against the offset
\(\Delta\log n_e\). The strongest
correlations with \(\Delta\log n_e\) are found for \(A_V\) (\(r_s=0.388\)),
\(N2\) (\(r_s=0.376\)), stellar mass (\(r_s=0.351\)), and
\(\log O_{32}\) (\(r_s=-0.300\)). Dust-corrected H$\alpha$ SFR also
correlates with the offset (\(r_s=0.261\)), whereas H$\alpha$ sSFR does not
(\(r_s=-0.034\), \(p=0.14\)). Thus, the same properties that generate the
color gradients in Figure~\ref{fig:relation} are primarily associated with the
difference between the [O II] and [S II] density diagnostics, not with the
absolute [S II]-based density itself. Larger offsets occur preferentially in
more massive, dustier, and more strongly low-ionization systems, whereas
high-ionization systems show smaller offsets.

Because these quantities are mutually correlated along the
mass--metallicity--dust--ionization sequence, we do not interpret any single
correlation coefficient as identifying an independent causal driver. The robust
conclusion is instead that the offset is tied to the
low-ionization/dust/metallicity sequence and is not explained by sSFR alone.

\section{Discussion and Conclusions}

Take advantage of the DESI emission-line catalog of \citet{zou24}, we show that the
[O II]-based densities are systematically higher than the [S II]-based densities
among star-forming galaxies. This is not a small-number result:
the full finite-density sample contains more than \(10^5\) galaxies, and even
the conservative high-precision calibration subset contains 1938 galaxies.

\subsection{Relation to previous work}

Our result is related to, but distinct from, several previous lines of work on
nebular electron-density diagnostics. High-resolution studies of nearby H II
regions provide the cleanest local benchmark. For example,
\citet{sanders2016} assembled 32 Galactic and extragalactic H II regions with
high-S/N, high-resolution spectra and found [O II]- and [S II]-based densities
to be broadly consistent with a one-to-one relation. By contrast, our DESI
spectra are galaxy-integrated: each spectrum mixes compact H II regions,
lower-ionization outer layers, diffuse ionized gas, and radial gradients in
metallicity, ionization parameter, and dust attenuation. The resulting
empirical [O II]--[S II] relation is therefore not an H II-region law, but an
integrated-galaxy calibration.

Our results are also related to, but distinct from, previous low- and
high-redshift galaxy studies. SDSS-quality integrated spectra have long
provided robust [S II]-based densities, but the [O II] doublet is only
marginally resolved in SDSS and therefore unsuitable for a precise
galaxy-by-galaxy comparison \citep{sanders2016}. High-redshift studies often
rely on one doublet at a time and have mainly focused on the evolution of the
absolute density scale \citep{kaasinen2017,kaasinen2018,davies2021,
reddy2023,topping2025}. The closest comparison at high redshift is the
GLASS-JWST analysis of \citet{li2025glass}, who found large scatter between
[O II]- and [S II]-based densities in the subset of galaxies with both
diagnostics available. Our DESI analysis extends this question to a large
low-redshift integrated-galaxy sample and shows that the offset has a clear
positive median and a systematic dependence on galaxy properties.

This emphasis of this work is also different from previous work on the absolute density
scale. Several studies have connected electron density or ionization conditions
to SFR, sSFR, compactness, or star-formation surface density
\citep[e.g.,][]{brinchmann2008,shimakawa2015,bian2016,kaasinen2017,
kaasinen2018,davies2021,reddy2023,topping2025}. Our control comparison in
Figure~\ref{fig:propertyfits} shows that, in the present low-redshift
high-precision DESI subset, most of these global properties are weak predictors
of the absolute [S II]-based density. The stronger trends appear instead when
the same properties are compared with the difference between the two density
diagnostics.

The most natural interpretation is that [O II] and [S II] weight different
low-ionization gas phases in galaxy-integrated spectra. Photoionization calculations show that [O II] and [S II] are emitted in
related but not identical low-ionization zones, and that their line ratios
respond differently to ionization parameter, metallicity, and density structure
\citep{osterbrock2006,kewley2001,kewley2019}. [S II] is emitted in
lower-ionization regions and is plausibly more affected by extended,
lower-density outer gas or diffuse ionized gas; spatially resolved studies
show that diffuse ionized gas enhances low-ionization lines such as [S II] and
[N II], thereby changing integrated nebular diagnostics
\citep{zhang2017,sanders2017,lacerda2018,belfiore2021}. In this context,
[O II] appears to be more strongly weighted toward denser low-ionization gas in
our integrated spectra. This picture is qualitatively supported by the
positive correlation of \(\Delta\log n_e\) with \(N2\), dust attenuation, and
stellar mass, and by its negative correlation with \(\log O_{32}\); the
connections among stellar mass, gas-phase metallicity, and dust attenuation in
star-forming galaxies make these coupled trends physically plausible
\citep{tremonti2004}. These trends are not easily explained as a simple scaling
of the absolute nebular density with SFR or sSFR alone; rather, they point to a
varying mixture of compact H II-region gas and extended low-ionization gas
across the galaxy population.

\subsection{Robustness and limitations}

\begin{figure*}
\centering
\includegraphics[width=0.98\textwidth]{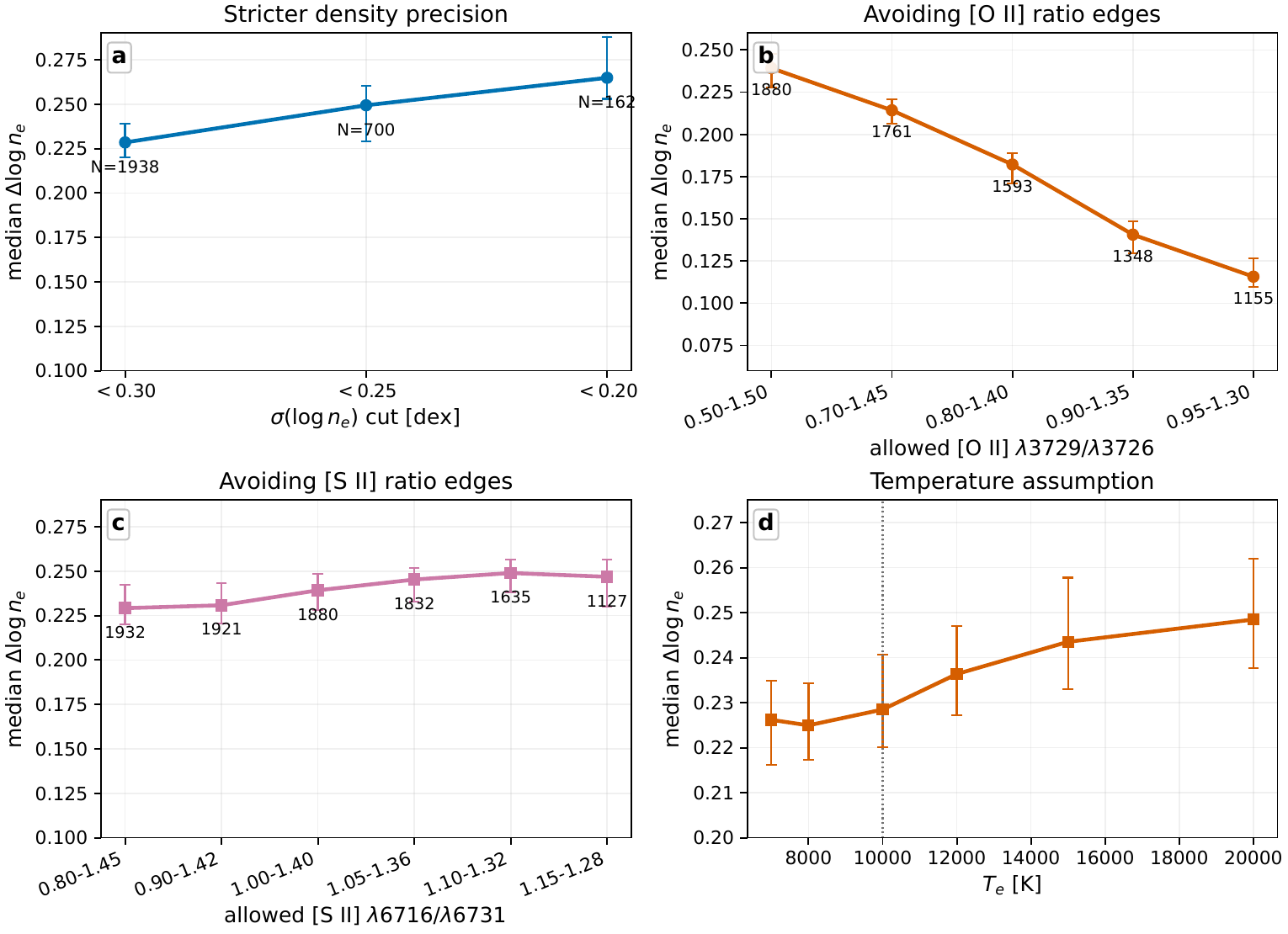}
\caption{Robustness of the [O II]--[S II] density offset. Panel (a) shows the
median offset for the fiducial and stricter density-precision cuts; labels give
the retained number of galaxies. Panel (b) shows the median offset after
increasingly conservative [O II] ratio cuts. Panel (c) repeats the exercise for
[S II] ratio cuts. Panel (d) shows the median offset after recomputing both
diagnostics at different fixed electron temperatures and reselecting the
\(\sigma(\log n_e)<0.3\) dex subset at each temperature.}
\label{fig:robust}
\end{figure*}

Figure~\ref{fig:robust} collects the main technical checks. The positive
[O II]--[S II] offset remains when the fiducial \(\sigma(\log n_e)<0.3\) dex
cut is tightened to 0.25 or 0.20 dex (panel~a), and it remains positive when the
[S II] doublet ratio is restricted away from its sensitivity edges (panel~c). The
[O II] ratio cut has a larger effect on the amplitude: removing progressively
more of the low-ratio, high-\(n_e({\rm OII})\) tail lowers the median offset
from about 0.24 dex to 0.12 dex (panel~b). This behavior is expected, because that cut
changes the density range over which the empirical relation is defined; it
does not erase the sign of the offset. The fixed-temperature assumption is
also not the source of the effect: recomputing both diagnostics at
\(T_e=7000\), 8000, 12000, 15000, and 20000 K, while reselecting the
\(\sigma(\log n_e)<0.3\) dex subset at each temperature, yields median offsets
in the range 0.226--0.248 dex (panel~d). Thus the
sign of the offset is robust, while its precise amplitude should be interpreted
as an empirical calibration over the stated line-ratio and sample-selection
range.

Several caveats remain. First, the calibration in
Equation~\ref{eq:calibration} is empirical and applies to DESI-like integrated
spectra selected as described above; it is not a universal H II-region law and
should not be applied indiscriminately to resolved H II regions or to
high-redshift galaxies with substantially different ISM conditions. Second,
the amplitude of the offset depends somewhat on how aggressively one excludes
[O II] ratios near the sensitivity limits. Third, although the temperature
tests show that the median offset is stable over \(T_e=7000\)–\(20000\) K,
object-by-object applications should use direct temperature estimates whenever
available. Fourth, the property--offset trends are based on integrated fiber
spectra and should ultimately be revisited with spatially resolved
spectroscopy. These caveats affect the detailed amplitude and physical
interpretation of the calibration, but not the basic empirical conclusion that
the two low-ionization doublets do not yield interchangeable electron-density
estimates in the selected DESI sample.

We conclude that [O II]- and [S II]-based electron-density diagnostics are not
interchangeable in integrated DESI spectra of low-redshift star-forming
galaxies. For the selected DESI sample, the median relation is approximately
$\log n_e({\rm OII})=0.75\,\log n_e({\rm SII})+0.83$.
This corresponds to \(n_e({\rm OII})\) exceeding \(n_e({\rm SII})\) by
roughly 0.1--0.3 dex over the commonly probed density range. Future studies
that compare electron densities across galaxy samples or redshift should
therefore account for this diagnostic offset explicitly, rather than assuming
that [O II] and [S II] are interchangeable low-ionization density tracers.

\begin{acknowledgments}

YR acknowledges supports from the CAS Pioneer Hundred Talents Program (Category B), the NSFC grants 12522302 and 12273037, and the USTC Research Funds of the Double First-Class Initiative. HZ acknowledges the supports from the National Natural Science Foundation of China (NSFC; grant Nos. 12120101003 and 12373010) and the Programs of National Astronomical Observatories Chinese Academy of Sciences with Grant Nos. E5ZQ7801 and E5ZB7801. This work is supported by the China Manned Space Program with grant no. CMS-CSST-2025-A06 and CMS-CSST-2025-A08.

\end{acknowledgments}

\end{document}